\documentclass[12pt]{article}
\newlength{\vshift}
\input epsf.tex
\newlength{\hshift}

\usepackage{amssymb}
\usepackage{amsmath}
\usepackage{amsthm}
\usepackage{amsopn}
\def\beq{\begin{equation}}
\def\eeq{\end{equation}}
\def\bea{\begin{eqnarray}}
\def\eea{\end{eqnarray}}

\begin{document}

 \vspace*{3cm}

\begin{center}

{ \Large \bf{ Short Distance Physics and Initial State Effects on the
CMB Power Spectrum } }

\vskip 4em

{ {\bf M.~Zarei} \footnote{e-mail: zarei@ph.iut.ac.ir }}

\vskip 1em

Department of Physics, Isfahan University of Technology, Isfahan
84156-83111, Iran
 \end{center}

 \vspace*{1.5cm}

\begin{abstract}
We investigate a modification in the action of inflaton due
to noncommutativity leads to a nonstandard initial vacuum and
oscillatory corrections in the initial power spectrum. We show
that the presence of these oscillations causes a drop in the WMAP
$\chi^{2}$ about $\Delta\chi^{2}\sim 8.5$. As a bonus, from the parameter estimation
done in this work, we show that the noncommutative parameters can
be precisely bound to $10^{16}$ GeV or $10^{4}$ GeV depending on the
inflation scale.
\end{abstract}
%%%%%%%%%%%%%%%%%%%%%%%%%%%%%%%%%%%%%%%%%%%%%%%%%%%%%%%%%%%%%%%%%%%%%%%%%%%%%%%%%%%%%%%%%%%%%%%%%%%%%%%%%%%%%%%%%%%%%%%%%%%
\newpage
\section{\large Introduction}
It has recently been emphasized that the effects of
trans-Planckian physics might be observable on cosmological scales
in the spectrum of Cosmic Microwave Background (CMB) radiation
\cite{brand1}-\cite{holman1}. An especially intriguing theory that
could makes this possible is inflation \cite{book1}-\cite{garcia}.
 The mechanism
of inflation  answers several questions that can not be solved in
the standard big bang cosmology. It is also the first predictive
theory for the origin of structures in the Universe and these
predictions have been verified to great accuracy using CMB
anisotropy experiments.

The phase of inflation must have lasted about 60 e-foldings so that the observed
structures in the Universe had enough time to be seeded from
vacuum fluctuations. Consequently it can be possible that
inflation is begun from scales with characterized wavelengths
much shorter than the Planck length $l_{P}$. Therefore one can
trace the trans-Planckian effects on the inflationary predictions.
For instance, due to trans-Planckian effects on the vacuum
fluctuations, the scale invariant primordial power spectrum
$P_{0}$ could be modified to a power spectrum which contains
superimposed oscillatory terms suppressed by the scale
$\sigma_{n}=(\frac{H}{\Lambda})^{n+1}
$ \cite{daniel1,niemeyer4,brand4,kaloper}. Here, $H$ is the Hubble
scale during inflation and $\Lambda$ is the scale associated with
the trans-Planckian physics.

Although the power spectrum and CMB
fluctuation can be influenced by the details of high energy scale
physics, there is no agreement on the size of these effects
and $n$ can be $n=0,1,2$ (see
\cite{daniel1,niemeyer4,brand4,kaloper}). For example using the
low energy effective field theory and general decoupling
arguments, Kaloper et al. \cite{kaloper} have shown a suppression
of the order of $\sigma_{1}=(\frac{H}{\Lambda})^{2}$ and have
discussed that it is too small to be observed during future CMB
experiments. This result strongly violates the first conclusions
that the deviation would be testable \cite{brand1,brand2}. The
argument of \cite{kaloper} has been criticized by some authors
\cite{brand5,weinberg}. Brandenberger and Martin in \cite{brand5}
have shown that instead of \cite{kaloper}, the trans-Planckian
physics can leave imprints on the CMB anisotropy. This approach is
the same as the Danielsson approach \cite{daniel1}. Both insist on
this fact that the vacuum state must be modified due to
trans-Planckian physics. Although the preliminary data analysis
\cite{WMAP1,WMAP2,WMAP3} showed that these oscillatory
modifications decrease the WMAP $\chi^{2}$ but recently
Groeneboom and Elgaroy, by investigating the Danielsson conclusion
\cite{daniel1}, have discussed that there is no significant
evidence for the simulated data to prefer the trans-Planckian
models \cite{WMAP11}. We have also analyzed the Danielsson formula
and observed no improvement in the $\chi^{2}$.

In this work we change the kinetic part of an
inflaton action by assuming a harmonic oscillatory term which
comes from noncommutativity to resolve the UV/IR mixing problem.
This term plays the role of a barrier potential and makes the vacuum
state to be a combination of negative and positive modes. We show
that such modification of a vacuum state leads to a power spectrum
containing oscillatory corrections similar to the Danielsson
result but with an important difference. Using the cosmological
Monte Carlo (CosmoMC) code,  we will show that due to the presence
of this difference, our model gives a drop in the $\chi^{2}$ in
spite of the Groeneboom and Elgaroy's negative conclusion. Having
parameters estimated by CosmoMC, the noncommutativity scale
$\Lambda_{NC}$ is determined as a function of inflation scale $H$.
Knowing a variety of bounds on $H$, we find the scale of
noncommutativity to be of the order of $10^{16}$ GeV or of the
order of $10^{4}$ GeV depending on what scales for inflation are
used.

The paper is organized as follows. In sections 2 and 3 we
briefly review the standard calculation of the inflationary power
spectrum when the trans-Planckian effects are taken into account.
In section 4 we derive our modified formula for the power spectrum
when an extra term is added to the free part of inflaton action
due to noncommutativity. Comparison of our theoretical
predictions with the WMAP data is given in section 5. In this
section we bound the noncommutative scale using parameters
estimated by CosmoMC.

%%%%%%%%%%%%%%%%%%%%%%%%%%%%%%%%%%%%%%%%%%%%%%%%%%%%%%%%%%%%%%%%%%%%%%%%%
\section{\large Initial Vacuum State}
The authors of \cite{brand5} have suggested that the evolution of
the inflaton modes can be separated into three phases I, II and
III shown in Figure 1. During phase I, the physical wavelength
$\lambda$ of modes is smaller than the scale $\Lambda\approx l_{P}$ and
the effects of short distance physics are expected to be important.
During phase II, the physical wavelength is larger than the Planck
scale but smaller than the Hubble radius $H^{-1}$.
Since during phase II the background is time-dependent, there is a
nontrivial mixing between creation and annihilation operators at
different times\cite{book3}. This kind of mixing takes place via
Bogoliubov transformation which through it a mode
$v(\eta)$ must be a linear combination of positive and negative
frequency initial modes\cite{book3}

\beq
v(\eta)=\alpha_{k}v^{(in)}_{k}(\eta)+\beta_{k}v^{(in)\ast}_{k}(\eta)\label{equ}
\eeq where $\alpha_{k}$ and $\beta_{k}$ are the Bogoliubov
coefficients which satisfy the normalization condition

\beq |\alpha_{k}|^{2}-|\beta_{k}|^{2}=1\eeq Finally in region III,
modes cross the Hubble radius $H^{-1}$ and squeeze.
\begin{figure}\vspace{-3cm}
\centerline{\epsfysize=5in\epsfxsize=6in\epsffile{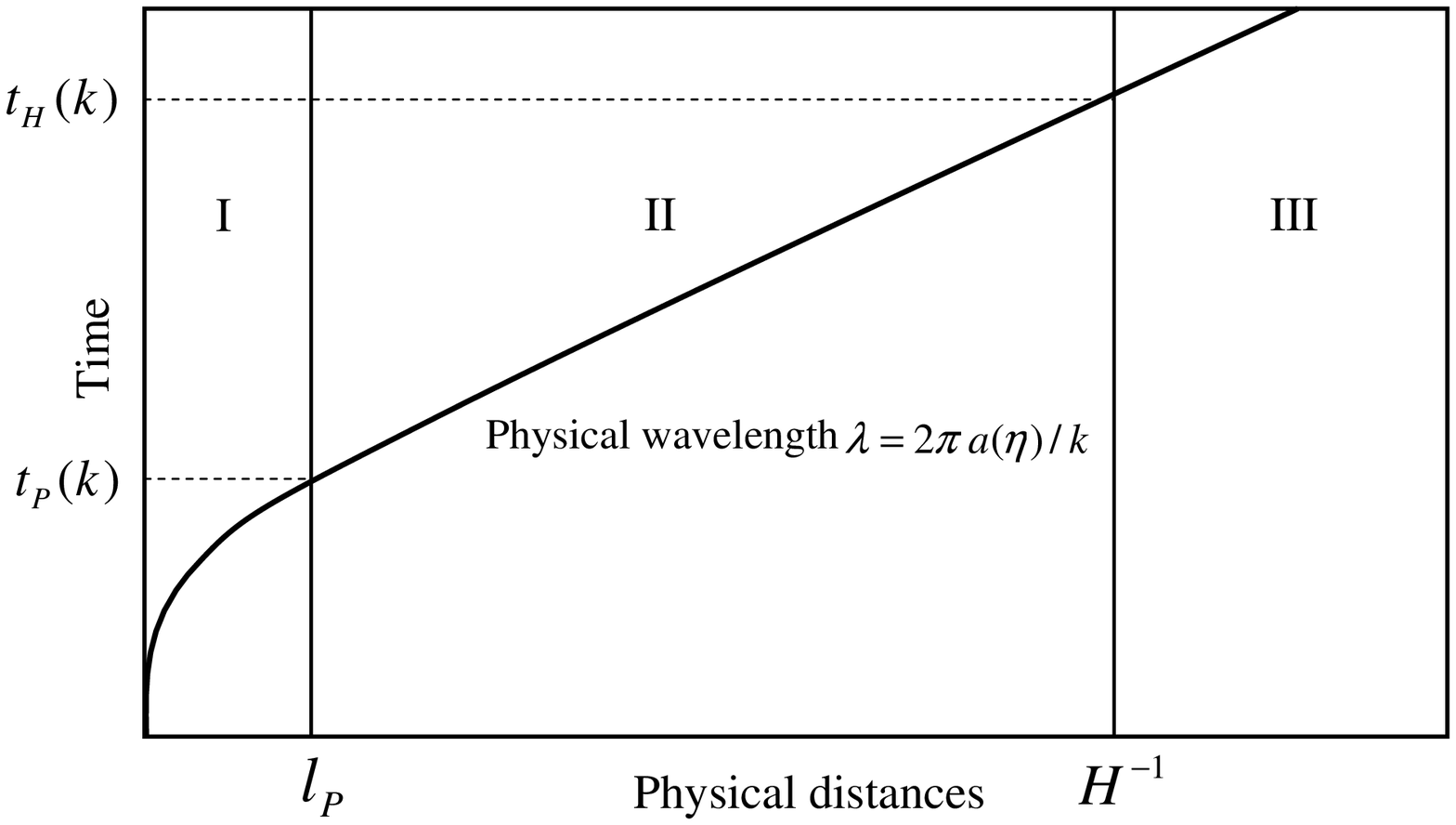}}\vspace{-3cm}
\caption{Periods
 }\label{fig0}
\end{figure}
The coefficients $\alpha_{k}$ and $\beta_{k}$ are determined by
matching (\ref{equ}) to the vacuum state. But in any expanding
universe such as inflationary period, the notion of a vacuum is ambiguous \cite{book3}. The reason is that $H$ is
time-dependent during inflation and so inflating space-time is not
exactly a de Sitter space-time in which defining vacuum state is
possible. This is a well-known problem in curved space-times
where the concept of the vacuum state is quite ambiguous due to the
absence of Killing vector fields \cite{book3}.

 A space-time such as
de Sitter space-time, admits a timelike Killing vector field,
then this provides a natural way to distinguish
positive and negative frequency modes and then similar to the
standard procedure in Minkowski space, associate these
distinguished modes with annihilation and creation operators. By
definition, the vacuum state of a de Sitter space-time, (known as the
Bunch-Davies vacuum \cite{BD}), will be the state that is
annihilated by all the annihilation operators. This vacuum state
is invariant under the symmetry group of the space-time. For inflationary theories although $H$ is time
dependent and consequently the Killing vector cannot be defined, this
ambiguity can be ignored by choosing the adiabatic vacuum when the
wavelength of a mode is much shorter than the curvature scale of
space-time. This vacuum is known as the adiabatic vacuum \cite{book3}.

In addition to this, there is also an extra ambiguity as
the result of trans-Planckian physics below the certain cut-off
$\Lambda$ through which the vacuum is ambiguous since the
physics of region I (Planck scales) is unknown. Danielsson approach \cite{daniel1} tries to solve
this kind of ambiguity.
 In this approach
there is no explicit assumption about the trans-Planckian physics
of region I and the emphasize is made only on the point that the
wavefunctions of the fluctuation modes are not in their vacuum
states when they enter into phase II \cite{daniel1}. Then for
 avoiding the ambiguity of region I which is the region of unknown physics,
one defines the vacuum at the time $\eta_{0}$ when the physical
momentum $p$ of a mode equals the scale of new physics i.e. $\Lambda$
\cite{daniel1}. The physical momentum $p$ and the comoving
momentum $k$ are related through

\beq k=ap=-\frac{p}{\eta H} \label{2}.\eeq
By imposing the initial conditions when $p=\Lambda$, we find the
conformal time $\eta_{0}$ as

 \beq \eta_{0}=-\frac{\Lambda}{Hk}.
\label{3}\eeq
Without knowledge of the physics beyond the scale $\Lambda$, in
order to choose the vacuum it is not necessary to take the limit
$\eta_{0}\rightarrow \infty$ but instead one had to stop at the value of
conformal time given by (\ref{3}). So the question about the
trans-Planckian physics is that of choosing the vacuum state
which clearly would be different from the adiabatic or Bunch-Davies
vacuum. An alternative vacuum leads to the corrections of order
$\sigma_{n}=(\frac{H}{\Lambda})^{n+1}$ in the power spectrum.

%%%%%%%%%%%%%%%%%%%%%%%%%%%%%%%%%%%%%%%%%%%%%%%%%%%%%%%%%%%%%%%%%%%%%%%%%
\section{\large Region II}

In this section we recall what happens in region II during
the inflation. During inflation
 the action of the massless scalar inflaton field $\varphi$
 minimally coupled to gravity is written as \cite{book1}-\cite{garcia}
 \beq S=\int\:d^{4}x\:\sqrt{-g}\left(\frac{1}{2}g^{\mu\nu}\partial_{\mu}\varphi\partial_{\nu}\varphi-V(\varphi)\right)
  \label{action}\eeq
where $g^{\mu\nu}$ is metric and $V(\varphi)$ is the inflaton potential. Usually one assumes a spatially flat, homogeneous and isotropic background
 with the metric
 \beq ds^{2}=a^{2}(\eta)\left(d\eta^{2}-d\textbf{x}\cdot d\textbf{x}\right)\label{metric}\eeq
where $\eta$ is the conformal time and $a(\eta)$ is the
scale factor. Here we restrict our attention to the scalar
perturbations in metric which will be gauge dependent. Hence the line element
in the longitudinal gauge, takes the form \beq
 ds^{2}=a^{2}(\eta)\left[(1+2\Phi(\eta,\textbf{x}))d\eta^{2}-(1-2\Psi(\eta,\textbf{x}))d\textbf{x}\cdot d\textbf{x}\right]\label{metric2}
 \eeq
where the space-time dependent functions $\Phi$ and $\Psi$ are
the two physical metric degrees of freedom which describe the scalar
metric fluctuations. In the absence of anisotropic stress, the two metric perturbation $\Phi$ and $\Psi$
coincide, $\Phi=\Psi$.
During inflation the inflaton field can be separated into a
classical background and a fluctuating part \beq
\varphi(\eta,\textbf{x})=\varphi_{0}(\eta)+\delta\varphi(\eta,\textbf{x})
\label{back}\eeq Because of the Einstein equation, the
metric fluctuation $\Phi$ is determined by the matter
 fluctuation $\delta\varphi$. Thus every thing can be
reduced to the study of a single gauge-invariant variable field
\cite{brand4,brand5}
 \beq
v=a\left(\delta\varphi+\frac{\varphi_{0}}{\mathcal{H}}\Phi\right)\equiv
z\mathcal{R},\label{v}\eeq where $a'=da/d\eta$,
$\mathcal{H}=\frac{a'}{a}$, $z\equiv
a\frac{\varphi'_{0}}{\mathcal{H}}$ and $\mathcal{R}$ denote the
curvature perturbation in comoving gauge \cite{brand4,brand5}. It
is convenient to work with the variable $\mu_{s}$ defined by
$\mu_{s}\equiv -\sqrt{2\kappa}v$ where $\kappa=8\pi/m_{P}$. After
quantizing the theory, in the Schr\"{o}dinger picture, we can write
$\mu_{s}(\eta,\textbf{x})$ as
 \beq
\mu_{s}(\eta,\textbf{x})=\frac{1}{(2\pi)^{\frac{3}{2}}}\int
d^{3}k\:\left[\:U^{II}_{k}(\eta)\:\hat{a}_{k}\:e^{i\textbf{k}\cdot
\textbf{x}}\:+\:U^{II\:\ast}_{-k}(\eta)\:\hat{a}^{\dag}_{-k}\:e^{-i\textbf{k}\cdot
\textbf{x}}\: \right], \label{mu}\eeq
 where $\hat{a}_{k}$ and
$\hat{a}^{\dag}_{-k}$ are respectively the creation and
annihilation operators. The evolution of $U^{II}_{k}(\eta)$ is
given by the Klein-Gordon equation of motion in Fourier space \beq
U^{II\:''}_{k}(\eta)+\left(k^{2}-
\frac{z''}{z}\right)U^{II}_{k}(\eta)=0. \label{motion}\eeq In
the case of slow-roll inflation limit the quantity $\frac{z''}{z}$
is reduced to \cite{brand4} \beq \frac{z''}{z}\sim
\frac{2+3\epsilon}{\eta^{2}},\label{z2}\eeq in which $\epsilon$ is the
slow roll parameter. The general solution for $U^{II}_{k}(\eta)$
is written in the form \beq U^{II}_{k}(\eta)=\frac{\sqrt{\pi}}{2}
\left|\eta\right|^{\frac{1}{2}}\left(\:\alpha_{k}\:e^{-\frac{i\pi}{2}(\frac{1}{2}-\nu)}\:H_{\nu}(|
k\eta|)+
\beta_{k}\:e^{\frac{i\pi}{2}(\frac{1}{2}-\nu)}\:H^{\ast}_{\nu}(|k\eta|)\:\right),\label{H}\eeq
where $H_{\nu}$ is the Henkel function of the first
kind. In the above expression, $\nu$ is a function of $\epsilon$ in the slow roll case.
The coefficients $\alpha_{k}$ and $\beta_{k}$ are the Bogoliubov coefficients
which are fixed by the initial condition. Using the asymptotic
form of the Hankel functions in the limit $k\eta\gg 1$\cite{book4} \beq
H^{(1)}_{\nu}(k\eta\gg 1)\approx
\sqrt{\frac{2}{\pi}|k\eta|}e^{i(k\eta-\pi\frac{\nu}{2}-\frac{\pi}{4})},
\eeq one can find the leading behavior of $U^{II}_{k}(\eta)$
 \beq
U^{II}_{k}(\eta)\approx
\frac{\alpha_{k}}{\sqrt{2k}}\:e^{-ik(\eta-\eta_{0})}+\frac{\beta_{k}}{\sqrt{2k}}\:e^{ik(\eta-\eta_{0})}+\cdot\cdot\cdot.
\label{u2}\eeq

Naively, one can choose a boundary initial condition by simply stating
that the modes in the limit $\eta\rightarrow -\infty$ are positive
frequency modes of the kind of Bunch-Davis vacuum which leads to
$\alpha=1$ and $\beta=0$. But as have been widely discussed
recently  \cite{brand1,brand2,brand3},  \cite{tanaka}-\cite{holman1},
by the trans-Planckian considerations, the vacuum state is not
a Bunch-Davis one. To solve this ambiguity, Danielsson
\cite{daniel1} assumed the following boundary condition \beq
\frac{d}{d\eta}\left(U^{II}_{k}/a\right)(\eta_{0})=-ik
\left(U^{II}_{k}/a\right)(\eta_{0})\label{bound}
 \eeq
 which gives the coefficients $\alpha$ and $\beta$ and consequently the
power spectrum $P(k)$ as \beq  P(k) = P_{0}(k)
\left( 1+\xi \left( \frac{k}{k_0}\right)^{-\epsilon} \sin
\left[\frac{2}{\xi } \left(\frac{k}{k_0} \right)^\epsilon + \phi
\right]\right),\label{dan} \eeq
where $\xi \equiv\frac{H}{\Lambda}$ and
$\phi$ is a phase factor. A similar result reported in \cite{pad} and also see \cite{cai1} and \cite{cai2} for another discussion and conclusion. In the following section we will drive
such a formula for trans-Planckian power spectrum by considering
an assumption about the physics in the region $I$ instead of
boundary condition (\ref{bound}). Our result for the power spectrum will be different from (\ref{dan}) and the difference is that
in our case two coefficients $\xi_{1}$ and $\xi_{2}$  appear instead of $\xi$.

%%%%%%%%%%%%%%%%%%%%%%%%%%%%%%%%%%%%%%%%%%%%%%%%%%%%%%%%%%%%%%%%%%%%%%%%%%%%%%%%%%%%%%%%%%%%%%%%%%%%%%%%%%%%%%%%%%%%%%%%
\section{\large Region I }

Since in
region I the complete theory of quantum gravity is still unknown, one way to tackle with it is to work with
effective scenarios. For instance, it is possible to include the
higher dimensional terms to the Lagrangian
\cite{kaloper,weinberg,shalm} or invoke the modified dispersion
relation models \cite{brand5}. Another possibility that has been greatly studied
in recent years is the noncommutativity of space-time \cite{brand3}.
Noncommutative space-time emerges in string
theory configuration  and is an effective picture of
the foamy space-time above the energy scale $\Lambda_{NC}$ (for review see \cite{2,3}). Field
theories on a noncommutative space-time

\beq \left[\:\hat{x}^{\mu},\:\hat{x}_{\nu}\:\right]=i\theta_{\mu\nu} \label{4}\eeq
are defined by replacing the ordinary product of fields by the
star product \cite{douglas,szabo} \beq
\phi_{1}\star\phi_{2}(x)=\int\frac{d^{d}kd^{d}q}{(2\pi)^{d}}
\tilde{\phi}_{1}(k)\tilde{\phi}_{2}(q)e^{ik_{\mu}\theta^{\mu\nu}q_{\nu}}e^{i(k+q)x}\;,
\label{5}\eeq where the $\theta_{\mu\nu}$ is a real antisymmetric
tensor which introduce a scale  $\Lambda_{NC}=\frac{1}{\sqrt{\theta}}$.
In Eq. (\ref{5}) the tilde denotes the Fourier transform. Because of this
star-product, a puzzling mixing of UV and IR scales known as UV/IR
mixing appeared during performing loop calculations in
noncommutative field theories \cite{Minwalla}. The
UV/IR mixing makes theories 
nonrenormalizable. Several attempts have been done in order to
solve this problem \cite{grosse,gurau,khodam}. For the case of
scalar theories, Grosse and Wulkenhaar in \cite{grosse}, have shown that one can
cancel out the UV/IR mixing problem by introducing a modification
in the free part of the action. They assume a harmonic
oscillatory term to the action such that the kinetic term becomes
\beq
S_{kin}[\phi]=\int\:d^{4}x\left(\:\frac{1}{2}\partial_{\mu}\phi\partial^{\mu}\phi+
\frac{\Omega^{2}}{2}\tilde{x}^{2}\phi^{2}\right), \label{6}\eeq where
$\Omega\in (\:0,1]$ and
$\tilde{x}_{\mu}=2\theta^{-1}_{\mu\nu}x^{\nu}$.
Furthermore in (\ref{6}), for the case of $\Omega^{2}=1$  the action is expected to enjoy a duality
transformation between position and momentum \cite{szabo2}
\begin{align}
\hat{\phi}(p) &\leftrightarrow \pi^2 \sqrt{|\det \theta|}\;\phi(x)\;, &
p_\mu &\leftrightarrow \tilde{x}_\mu\;,
\label{duality}
\end{align}
where $ \hat{\phi}(p_a)=\int d^4 x \;\mathrm{e}^{(-1)^a i
  p_{a,\mu} x_a^\mu} \phi(x_a)$.

Inspired by this
modification, we suggest that the kinetic term of an inflaton action must be modified
due to quantum properties of space-time on the scales where
trans-Planckian effects become important. Since the noncommutativity is one of effective pictures of the trans-Planckian physics, hence for the inflatons, a modified action such as (\ref{6}) is considered.
 Therefor we consider the
harmonic oscillator term $\frac{\Omega^{2}}{2}\varphi(x)
\tilde{x}^{2}\varphi(x)$ as a correction to the kinetic part of
the inflaton action (\ref{action}), in region I. The noncommutative parameter
$\theta_{\mu\nu}$ is  also restricted to the following special
case where as the result of it the equation of motion of the inflaton is simplified

\beq \theta_{\mu\nu}=\left(
                       \begin{array}{cccc}
                         0 & 0 & 0 & \theta \\
                         0 & 0 & \theta & 0 \\
                         0 & -\theta & 0 & 0 \\
                         -\theta & 0 & 0 & 0 \\
                       \end{array}
                     \right)
 \eeq
Consequently under this assumption, $\tilde{x}^{2}$ is given by

\beq \tilde{x}^{2}=
\left(2\:\theta^{-1}_{\mu\nu}x^{\nu}\right)^{2}=4\frac{a^{2}}{\theta^{2}}
\left[(x^{3})^{2}-(x^{1})^{2}-(x^{2})^{2}-(\eta)^{2}\right]
  \label{x}\eeq
and the kinetic part of the inflaton action is written as
 \beq S_{\mathrm{kin}}=\int\:d^{4}x\:\sqrt{-g}\left(\frac{1}{2}\:g^{\mu\nu}\partial_{\mu}\varphi\:\partial_{\nu}\varphi\:+\:2\:\frac{a^{2}}{\theta^{2}}
\left[(x^{3})^{2}-(x^{1})^{2}-(x^{2})^{2}-(\eta)^{2}\right]\varphi^{2}\right).
  \label{action2}\eeq

Now the variable $\mu_{s}(\eta,\textbf{x})$ is decomposed into the
Fourier modes in the following general form
\beq \mu_{s}(\eta,\textbf{x})=\frac{1}{(2\pi)^{\frac{3}{2}}}
\int[d\textbf{k}]\big(\:\hat{a}_{k}\:U_{k}(\eta)\:V_{k}(\textbf{x})\:+
\:\hat{a}^{\dag}_{k}\:U^{\ast}_{k}\:V^{\ast}_{k}(\textbf{x})\:\big)\;,\label{ex}\eeq
where $[d\textbf{k}]$ is the measure of the integral and the $V_{k}(\textbf{x})$ is the
eigenfunction of the three dimensional Laplace operator
$\Delta_{3}$
\bea \:\left(\Delta_{3}+(x^{3})^{2}-(x^{2})^{2}-(x^{1})^{2}
\:\right)V_{k}(\textbf{x})\!\!\!\!\!\!\! &&=\frac{1}{\sqrt{-g}}\:
\partial_{i}\left(\sqrt{-g}g^{ij}\partial_{j}\right)V_{k}(\textbf{x})\nonumber\\&&
\:\:\:\:\:+\left((x^{3})^{2}-(x^{2})^{2}-(x^{1})^{2}\right)
\:V_{k}(\textbf{x})\nonumber\\&& =k^{2}V_{k}(\textbf{x})\;,
\label{VV}\eea in which the metric $g^{ij}$ has been defined in
(\ref{metric}) and $k^{2}$ is the eigenvalue. Also in the
expansion (\ref{ex}), $U_{k}(\eta)$ obeys the following equation

\beq
U''_{k}(\eta)+\left[\left(k^{2}-\frac{a''}{a}\right)+\frac{2}{\theta^{2}}\eta^{2}\right]U_{k}(\eta)=0.
\label{uu}\eeq In the case of the absence of the second term in
equation (\ref{VV}), the $V_{k}(\textbf{x})$ are simply plane-wave
$e^{i\textbf{k}\cdot \textbf{x}}$ and the integration measure
in (\ref{ex}) for this case is $\int[d\textbf{k}]=\int d^{3}k$.
The differential equation (\ref{VV}) can be solved by separating
$V_{k}(\textbf{x})$ into \beq
V_{k}(\textbf{x})=V_{k_{3}}(x_{3})V_{k_{1},k_{2}}(x_{1},x_{2})
\eeq The differential equation for the $V_{k_{3}}(x_{3})$
reduces to the harmonic oscillator differential equation, which its
solution is given in terms of Hermite polynomials. Thus, the
wave number $k_{3}$ takes discrete values which we label by
positive integers $k\geq 1$. Also for the
$V_{k_{1},k_{2}}(x_{1},x_{2})$ we find a differential equation,
which its solution is known as parabolic cylinder functions
\cite{book4}. The parabolic cylinder function, $D_{p}(z)$ with continues $p$, is a class of special functions defined as the solution
to the following differential equation
\beq \frac{d^{2}D_p(z)}{dz^{2}}+\left(\:z^{2}+\lambda(p)\:\right)D_p(z)=0, \eeq
together with the integral representation as
\beq D_{p}(z)=\frac{e^{\frac{-z^{2}}{4}}}{\Gamma(-p)}\:\int_0^\infty e^{-xz-\frac{x^{2}}{2}}x^{-p-1}dx\:\:\:\:\:\:\:\:\:\:\:\:(\mathrm{Re\:p<0}). \eeq
Since $k_3$ is discrete and $k_{1}$ and $k_{2}$ are continuous,
the integration measure in (\ref{ex}) is
given by \beq
\int[d\textbf{k}]=\int_{0}^{\infty}\int_{0}^{\infty}
d^{3}k_{1}d^{3}k_{2}\sum_{k_{3}=1}^{\infty}. \eeq
Because the modes are in 
region I, one expects $\frac{a''}{a}\rightarrow 0$ for the
time-dependent part of modes in (\ref{uu}). Hence equation (\ref{uu})
is simplified to \beq
U''_{\kappa}(\tau)+\big(\kappa^{2}+\tau^{2}\big)U_{\kappa}(\tau)=0\;,
\label{time} \eeq in which  $U_{k}(\eta)$, $\eta$
and $k$ have been redefined as follows,

\beq
U_{\kappa}(\tau)=\left(\frac{\sqrt{2}}{\theta}\right)^{\frac{1}{4}}U_{k}(\eta),\:\:\:\:\:\:\:\:\:\:\:
\tau=\frac{(2)^{\frac{1}{4}}}{\sqrt{\theta}}\eta,
\:\:\:\:\:\:\:\:\:\:\: \kappa^{2}=\sqrt{2}\theta k^{2}. \eeq The
time evolution of modes in (\ref{time}) is equivalent to the
following situation. Consider a scalar inflaton on a non-static
background \beq ds^{2}=C(\eta)\big(d\eta^{2}-d\textbf{x}\cdot
d\textbf{x}\big). \eeq It can be shown that with a conformal scale
factor representing a bouncing universe

\beq C(\eta)=a^{2}+b^{2}\eta^{2} \:\:\:\:\:\:\:\:
-\infty<\eta<\infty,\eeq one obtains the same equation for the evolution of
time-dependent modes \cite{book3}. The exact solution
to the equation (\ref{time}) is then given in terms of parabolic
cylinder functions \cite{book4}

\beq U^I_{\kappa}(\tau)\equiv U_{\kappa}(\tau)=D_{-\frac{1+i\kappa^{2}}{2}}(\:(1+i)\tau\:).
\label {solo}\eeq

Now we study the asymptotic behavior of
$U^{I}_{\kappa}(\tau)\equiv U_{\kappa}(\tau)$ modes living in
region I in order to match them with $U^{II}_{\kappa}(\tau)$ modes living in region II at
the boundary $\eta_{0}$. This helps us to obtain
coefficients $\alpha_{k}$ and $\beta_{k}$. The parabolic cylinder
function $D_{p}(z)$ has the following linear relation \cite{book4}

\beq D_{p}(z)=e^{-p\pi
i}\:D_{p}(-z)\:+\:\frac{\sqrt{2\pi}}{\Gamma(-p)}\:e^{-\frac{(p+1)\pi
i}{2}}\:D_{-p-1}(iz)\;, \label{linear}\eeq where $z=(1+i)\tau$ and
$p=-\frac{1+i\kappa^{2}}{2}$. This relation is a useful relation
in order to rewrite the modes $U^{I}_{\kappa}(\tau)$ as a
combination of positive and negative modes. The asymptotic
behavior of $D_{p}(z)$ for the large values of $p$ is \cite{book4}
\beq D_{p}(z)\approx
\frac{2^{\frac{-1}{4}+\frac{p}{2}}}{\left(8(-\frac{1}{4}-\frac{p}{2})\right)^{\frac{1}{4}}}
\:\exp\left(\left(\frac{1}{4}+\frac{p}{2}\right)\ln
\left(\frac{1}{4}-\frac{p}{2}\right)-2\sqrt{-\left(\frac{1}{4}+\frac{p}{2}\right)\frac{z^{2}}{2}}\:\right).
\label{expan}\eeq Substituting (\ref{expan}) into the relation
(\ref{linear}) with $p=-\frac{1+i\kappa^{2}}{2}$, we find \bea
D_{-\frac{1+i\kappa^{2}}{2}}((1+i)\tau)\approx &&
\!\!\!\!\!\!\!\!
\frac{2^{-\frac{i\kappa^{2}}{4}}\:e^{\left(\frac{1}{2}+\frac{i\kappa^{2}}{4}\right)\pi
i-i\frac{\kappa^{2}}{4}\ln\frac{i\kappa^{2}}{4}}}{\left(2e^{\frac{i\pi}{2}}\kappa^{2}\right)^{\frac{1}{4}}}\:
e^{-i\kappa\tau} \nonumber\\&&
\!\!\!\!\!\!\!\!\!\!+\frac{\sqrt{2\pi}}{\Gamma\left(\frac{1}{2}+i\frac{\kappa^{2}}{2}\right)}
\frac{2^{\frac{i\kappa^{2}}{4}}\:e^{\left(\frac{1}{2}-\frac{i\kappa^{2}}{4}\right)\frac{-\pi
i}{2}+i\frac{\kappa^{2}}{4}\ln\frac{-i\kappa^{2}}{4}}}{\left(2e^{\frac{-i\pi}{2}}\kappa^{2}\right)^{\frac{1}{4}}}\:
e^{i\kappa\tau}
\nonumber\\&&\!\!\!\!\!\!\!\!\!\!\!\!\!
\approx\frac{i}{(2\kappa^{2})^{\frac{1}{4}}}\exp\left(-\frac{i\pi}{8}-\frac{\kappa^{2}\pi}{2}
\right)\:e^{-i\kappa \tau}\nonumber\\&&\!\!\!\!\!\!\!\!\!\!-
\frac{i\sqrt{2\pi}}{\Gamma\left(\frac{1}{2}+i\frac{\kappa^{2}}{2}\right)(2\kappa^{2})^{\frac{1}{4}}}
\exp\left(\frac{i3\pi}{8}-\frac{\kappa^{2}\pi}{4}\right)\:e^{i\kappa \tau}. \:
  \label{u} \eea
Thus $U^{I}_{k}(\eta)$ can be written as \beq
U^{I}_{k}(\eta)\approx\frac{1}{\sqrt{2k}}\:\alpha_{k}\:e^{-ik\eta}+
\frac{1}{\sqrt{2k}}\:\beta_{k}\:e^{ik\eta}, \label{u1}\eeq where
the coefficients $\alpha_{k}$ and $\beta_{k}$ are found by
matching the solution (\ref{u}) with  $U^{II}_{k}(\eta)$ in
(\ref{u2}) at $\eta=\eta_{0}$ as\beq
\alpha_{k}=ie^{-\frac{i\pi}{8}-\frac{k^{2}\pi
\theta}{2\sqrt{2}}}e^{ik\eta_{0}},\label{alpha} \eeq and \beq
\beta_{k}= \frac{-i\sqrt{2\pi}}{\Gamma\left(\frac{1}{2}+i\frac{\theta
k^{2}}{2\sqrt{2}}\right)}\:e^{-\frac{3\pi i}{8}-\frac{k^{2}\pi
\theta}{4\sqrt{2}}}e^{-ik\eta_{0}}.\label{beta}\eeq Using the
identify \cite{book4} \beq
\left|\Gamma\left(\frac{1}{2}+iy\right)\right|^{2}=\frac{\pi}{\cosh \pi y} \eeq one can
verify the normalization relation \beq |\alpha|^{2}-|\beta|^{2}=1\label{norm}
\eeq In general the gamma function can be represented in the
following form \beq \Gamma\left(\frac{1}{2}+ik^{2}\theta\right)=
\left|\Gamma\left(\frac{1}{2}+ik^{2}\theta\right)\right|\:e^{i\varphi(k^{2}\theta)}=\frac{\sqrt{\pi}}{\cosh^{1/2}
\pi k^{2}\theta }\:e^{i\varphi(k^{2}\theta)}\;, \label{g}\eeq where
$\varphi(k^{2}\theta )$ is a phase factor.  We will employ this
relation for the next calculations of $P(k)$.

 Using equations (\ref{alpha}) and (\ref{beta}), one can calculate the corresponding energy density and pressure by the mean value of energy-momentum tensor $\left<T_{\mu\nu}\right>$  in the trans-Planckian region \cite{energy1}
\bea \left<\rho\right>=\frac{1}{4\pi^{2}a^{4}}\int dkk^{2}\left(a^{2}\left|\left(\frac{U_k^I}{a}\right)'\right|^{2}+k^{2}\left|U_k^I\right|^{2}\right)
 \\ \left<p\right>=\frac{1}{4\pi^{2}a^{4}}\int dkk^{2}\left(a^{2}\left|\left(\frac{U_k^I}{a}\right)'\right|^{2}-\frac{k^{2}}{3}\left|U_k^I\right|^{2}\right) \eea
Inserting the mode function $U_k^I(\eta)$ into the vacuum expressions of the energy density and pressure and using the normalization condition (\ref{norm}), one finds
\bea \left<\rho\right>=\frac{1}{4\pi^{2}a^{4}}\int dkk^{3}\left|\beta_k\right|^2
 \\ \left<p\right>=\frac{-1}{4\pi^{2}a^{4}}\int dkk^{3}\left(\frac{2}{3}+\frac{1}{3}\left|\beta_k\right|^2\right) \eea
in the trans-Planckian region.
 Now at this stage we turn to the superhorizon scales ($k\eta\ll 1$) and start to
compute the power spectrum. First we substitute (\ref{alpha})
and (\ref{beta}) in (\ref{H}) and take  the superhorizon limit.
Since in this limit the Hankel functions take the following form
\beq H_{\nu}(k\eta\ll 1)\approx
\sqrt{\frac{2}{\pi}}\:e^{-i\frac{\pi}{2}}2^{\nu-\frac{3}{2}}\frac{\Gamma(\nu)}{\Gamma(\frac{3}{2})}
(k\eta)^{-\nu} \eeq The amplitude of $U^{II}_{k}$ modes is given
by \beq |U^{II}_{k}|\approx
\frac{2^{\nu-\frac{3}{2}}}{\sqrt{2k}}\frac{\Gamma(\nu)}{\Gamma(\frac{3}{2})}\:|\alpha_{k}+\beta_{k}|(k\eta)^{-\nu}
\eeq which gives the scalar power spectrum as \bea
P(k)\!\!\!\!\!\!\!\!\!&&=\frac{k^{3}}{2\pi^{2}}\left|\frac{U^{II}_{k}}{a}\right|^{2}
=P_{0}(k)\:|\alpha_{k}|^{2}\left(1+2Re\left(\frac{-\beta}{\alpha}\right)+\frac{|\beta|^{2}}{|\alpha|^{2}}\right)\;,\label{po}
\eea where $P_{0}$ is the ordinary power spectrum for the scalar
modes given by \cite{brand5} \beq P_{0}\approx
\left(\frac{H}{2\pi}\right)^{2}\left(\frac{k}{aH}\right)^{3-2\nu}. \eeq Usually $P_0$ is
parameterized by \beq
P_{0}=A_{s}\left(\frac{k}{k_{\star}}\right)^{n_{s}-1}, \eeq in which
$n_{s}$ is the conventional definition of spectral index
and $A_{s}$ is the spectral amplitude for the scalar perturbation.
$k_{\star}$ is a scale which is fixed to be $0.05\: Mpc^{-1}$
\cite{lewis1}. For the case of slow roll inflation, the Hubble
parameter $H$ is $k$-dependent with the form \cite{daniel2} \beq
\frac{H}{H_{0}}\sim \left(\frac{k}{k_\star}\right)^{-\varepsilon}, \eeq where
$\varepsilon$ is slow-roll parameter. Then the quantity $k\eta$
varies as
 \beq
k\eta=\frac{\Lambda_{NC}}{H}\sim
\frac{\Lambda_{NC}}{H_{0}}\:\left(\frac{k}{k_{\star}}\right)^{\varepsilon}.\label{eps}
\eeq Similarly, the quantity $k^{2}\theta$ varies as \beq
k^{2}\theta=\frac{a^{2}p^{2}}{\Lambda_{NC}}=\left[\:\frac{\Lambda_{NC}}{H_{0}}
\:\left(\frac{k}{k_{\star}}\right)^{-\varepsilon}\label{k}\:\right]^{2}.
\eeq We now substitute (\ref{alpha}) and (\ref{beta}) in
(\ref{po}) and use (\ref{eps}), (\ref{k}) and (\ref{g}) to find
$P(k)$ as
\bea
P(k)&&\!\!\!\!\!\!\!\!= A_{s}\left(\frac{k}{k_{\star}}\right)^{n_{s}-1}\nonumber \\&&\!\!\!\!\!\!\!\!\times
\left(\:e^{-\frac{k^{2}\theta\pi}{\sqrt{2}}}
+\sqrt{2}\:e^{-\frac{3}{4\sqrt{2}}k^{2}\theta\pi}\:\cosh^{1/2}\left(\frac{k^{2}\theta\pi}{2\sqrt{2}}\right)\cos(\pi/4+2k\eta)+1+
e^{-\frac{k^{2}\theta\pi}{\sqrt{2}}} \right) \eea where \beq
e^{-c\:k^{2}\theta}=e^{-c\:\left(\frac{\Lambda_{NC}}{H_{0}}\right)^{2}
\:\left(\frac{k}{k_{\star}}\right)^{-2\varepsilon}}.\eeq
 Since the $k^{2}\theta$ is large in value, the
exponential terms decay very fast. Thus it is possible to
approximate the exponentials with a polynomial function as follows

\bea
e^{-c\:k^{2}\theta}&&\!\!\!\!\!\!\!\!=e^{-c\:\left(\frac{\Lambda_{NC}}{H_{0}}\right)^{2}\:\left(\frac{k}{k_{\star}}\right)^{-2\varepsilon}}
\nonumber \\&&\!\!\!\!\!\!\!=c_{1}\times\left(\frac{H_{0}}{\Lambda_{NC}}\right)^{2}\left(\frac{k}{k_{\star}}\right)^{-2\varepsilon}+
\:c_{2}\times\left(\frac{H_{0}}{\Lambda_{NC}}\right)^{4}\left(\frac{k}{k_{\star}}\right)^{-4\varepsilon}+\cdot\cdot\cdot\:\:.
\eea
Keeping only terms up to order $\frac{H_{0}}{\Lambda_{NC}}$ and
redefining the coefficient $c_{1}\frac{H_{0}}{\Lambda_{NC}}$ as a
new parameter $\xi_{1}$ we will find  \beq
P(k)=A_{s}\left(\frac{k}{k_{\star}}\right)^{n_{s}-1}\times\left(1+
\:\xi_{1}\:\big(\frac{k}{k_{\star}}\big)^{-\varepsilon}\cos\left(\:\pi/4+\frac{2}{\xi_{2}}
\:\big(\frac{k}{k_{\star}}\big)^{\varepsilon}\right) +\cdot\cdot\cdot
\right), \label{pow} \eeq where
$\xi_{2}=\sigma_{0}=\frac{H_{0}}{\Lambda_{NC}}$. This result that
the coefficients $\xi_{1}$ and $\xi_{2}$ in the power spectrum formula
are not equal, is the
difference between our result and Danielsson formula and will play
a crucial role in CMB data analysis.

We saw that by considering
noncommutativity corrections to the free part of the action
of inflatons in region $I$, the power
spectrum receives  oscillation corrections.  This result has a
simple physical interpretation. With a suitable changing of
variables,
 equation (\ref{time}) for the
primordial modes is similar to the
 Schr\"{o}dinger equation for a wave function $\psi(x)$
of a quantum mechanical particle in a one dimensional barrier
potential $V(x)=x^{2}$

\beq \frac{d^{2}\psi}{dx^{2}}+(E-V(x))\psi=0. \label{V}\eeq Because of
this kind of barrier potential we expect modes in region I be
a superposition of the incoming modes and the reflected modes which have
been scattered off the barrier.
%%%%%%%%%%%%%%%%%%%%%%%%%%%%%%%%%%%%%%%%%%%%%%%%%%%%%%%%%%%%%%%%%%%%%%%%%%%%%%%%%%%%%%%%%%%%%%%%%%%%%%%%%%%%%
%%%%%%%%%%%%%%%%%%%%%%%%%%%%%%%%%%%%%%%%%%%%%%%%%%%%%%%%%%%%%%%%%%%%%%%%%%%%%%%%%%%%%%%%%%%%%%%%%%%%%%%%%%%%%
\section{\large Effects on CMB Temperature Fluctuations}
\begin{figure}\vspace{-5cm}
\centerline{\epsfysize=7in\epsfxsize=6in\epsffile{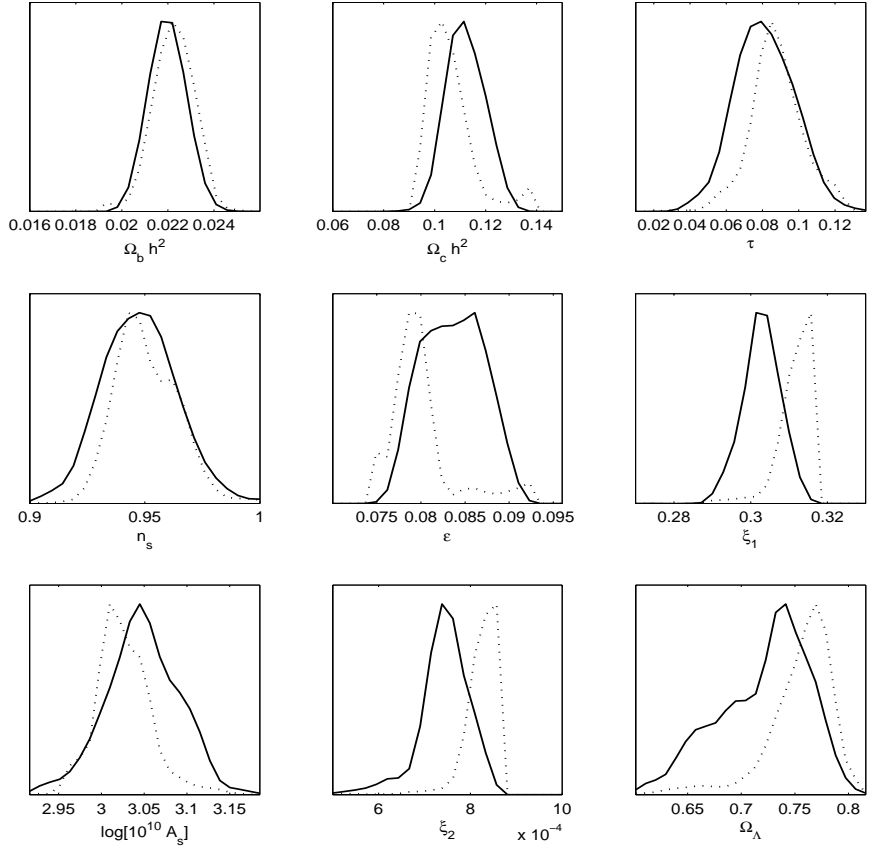}}\vspace{0.5cm}
\caption{One dimensional marginalized posterior probability
distribution (solid lines) and mean likelihood (dotted lines) for
the cosmological parameters.
 }\label{fig1}
\end{figure}

\begin{figure}\vspace{-4cm}
\centerline{\epsfysize=7in\epsfxsize=6in\epsffile{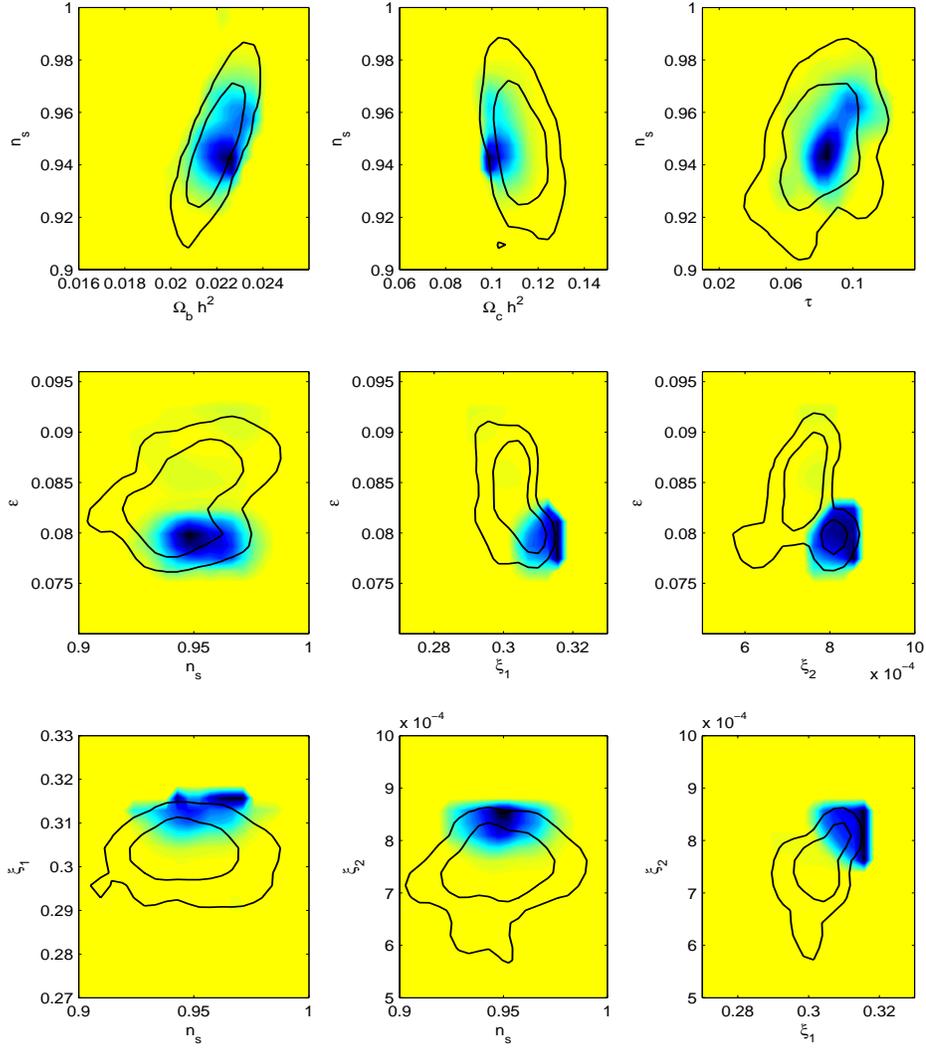}}\vspace{0.5cm}
\caption{Posterior constraint for our model. The contours show the
68 \% and 95 \% confidence limits from the marginalized
distribution.
 }\label{fig2}
\end{figure}

\begin{figure}
\centerline{\epsfysize=4in\epsfxsize=5in\epsffile{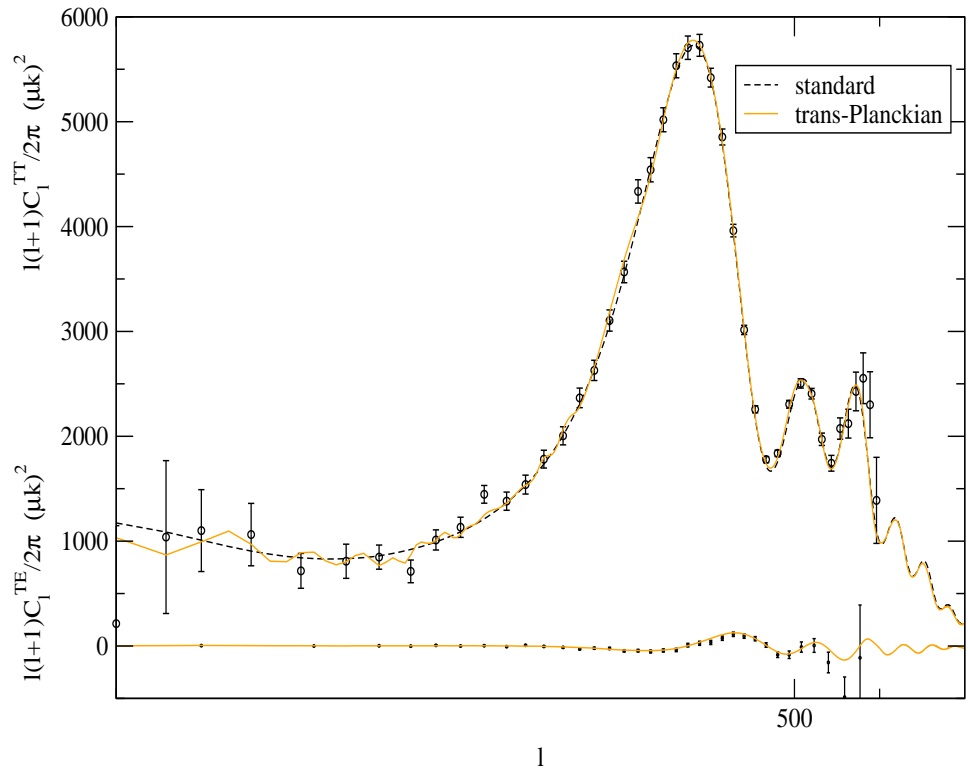}}
\caption{The best fit obtained for our trans-Planckian  angular TT
(up) and TE (down) power spectrum  compared to the best fit
standard inflationary power spectrum and five years WMAP data.
 }\label{fig3}
\end{figure}

\begin{figure}
\centerline{\epsfysize=4in\epsfxsize=5in\epsffile{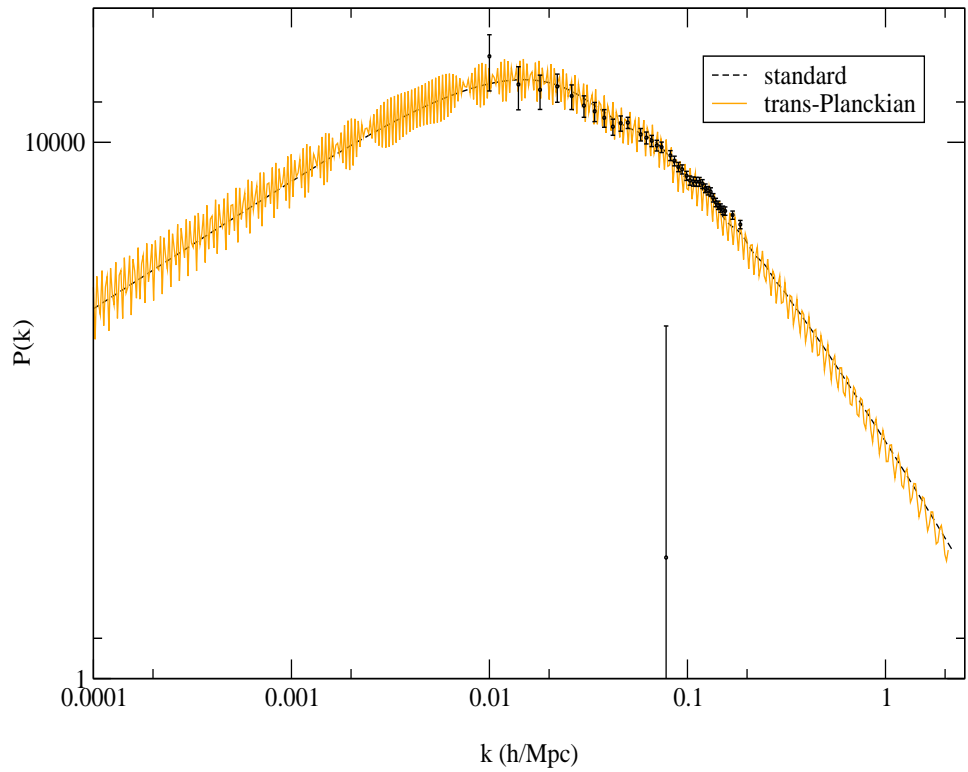}}
\caption{ Predicted matter power spectrum for our trans-Planckian
model compared to the 2dF data\cite{2dF}.
 }\label{fig5}
\end{figure}

In this section a CMB data analysis is provided in order to show that the agreement of
 our modified power spectrum with the CMB data is better than the standard one.
The fluctuations in the temperature of CMB radiation can be
expanded in spherical harmonics \cite{book1,book2}

\beq \frac{\Delta
T(\theta,\phi)}{T}=\sum_{l=2}^{\infty}\sum_{m=-l}^{m=l}
\:a_{lm}Y_{lm}(\theta,\phi)\eeq
for convenience, we have excluded the monopole
and dipole terms. The initial power spectrum $P(k)$ given in
(\ref{pow}) is related to the CMB anisotropy through the angular
power spectrum $C_{l}$ which is defined by two-point correlation
function of the temperature fluctuation

\beq \left< \frac{\Delta T}{T}(\hat{n}_{1})\:\frac{\Delta
T}{T}(\hat{n}_{2})
\right>\:=\sum_{l=2}^{\infty}\:\frac{2l+1}{4\pi}\:C_{l}
\:P_{l}(\cos\:\theta), \eeq where $P_{l}(cos\:\theta)$ is the
Legendre polynomials  and
$cos\:\theta=\hat{n}_{1}\cdot\hat{n}_{2}$. The $\hat{n}_{1}$ and
$\hat{n}_{2}$ are unit vectors pointing to arbitrary direction on
the sky. The angular power spectrum $C_{l}$ can be related to
$P(k)$ through

\beq C_{l}=4\pi \int_{0}^{\infty} \:T^{2}(k)P(k)\:\frac{dk}{k}\eeq
in which $T(k)$ is the transfer function. Since Thomson scattering
polarizes light \cite{hu}, there are also angular power spectrum
coming from the polarization. The polarization can be divided into
a curl (B) and curl-free (E) component which yields four
independent angular power spectrums as $C^{TT}_{l}$,$C^{EE}_{l}$,
$C^{BB}_{l}$ and the $T-E$ cross correlation $C^{TE}_{l}$ \cite{hu}. The
WMAP experiment has reported data only on $C^{TT}_{l}$ and
$C^{TE}_{l}$ \cite{hu}. The previous data analysis has reported
different conclusions about the signatures of trans-Planckian
modification of power spectrum
\cite{WMAP1}-\cite{WMAP11} and \cite{WMAP4}-\cite{WMAP10}.
The analysis of \cite{WMAP1,WMAP2,WMAP3} shows that the presence
of oscillations in power spectrum causes an important drop in the
WMAP $\chi^{2}$ of about $\Delta\chi^{2}\approx 10$.
$\chi^{2}$ is defined as $\chi^{2}=-2\log\mathcal{L}(x|p)$ in
which $\mathcal{L}(x|p)$ denotes the likelihood which is a conditional
probability function that allows us to estimate unknown
cosmological parameters $p$ based on CMB data $x$ and satisfies the
normalization condition $\int\mathcal{L}(x|p)dx=1$.
In spite of
the conclusion of \cite{WMAP1,WMAP2,WMAP3}, the recent analysis of
\cite{WMAP11} claims no evidence for trans-Planckian oscillations
especially in the Danielsson model. Now we are going to verify our prediction for the
trans-Planckian power spectrum by analyzing the WMAP data.
In the previous section we found trans-Planckian oscillatory
corrections to the power spectrum (\ref{pow}) in which the coefficients
$\xi_{1}$ and $\xi_{2}$ are not equal in comparison to the
Danielsson formula \cite{daniel1}.
 In order to compare our
result (\ref{pow}) with the recent WMAP data, we present a
Bayesian model selection analysis  data using the CosmoMC
(Cosmological Monte Carlo) code developed in \cite{lewis1} which
makes use of the CAMB program \cite{lewis2}. This program employs a
Markov-Chain Monte Carlo (MCMC) sampling procedure to explore the
posterior distribution. A Bayesian analysis provides a coherent
approach to estimating the values of the parameters, $p$, and their
errors and a method for determining which model, $M$, best describes
the data $x$. Bayes theorem states that \beq
P(p|x,M)=\frac{P(x|p,M)P(p|M)}{P(x|M)}, \eeq where $P(p|x,M)$ is
the posterior, $P(x|p,M)$ is the likelihood, $P(p|M)$ is the prior and
$P(x|M)$ is the Bayesian evidence. Conventionally, the result of a
Bayesian parameter estimation is the posterior probability
distribution given by the product of the likelihood and prior. The
parameter space we consider is $h$ (the dimensionless Hubble
constant), $\Omega_{b}$ (the amount of baryons), $\Omega_{c}$ (the
amount of cold dark matter), $\Omega_{\Lambda}$ (the amount of
dark energy), $\tau $ (the redshift of reionization), $n_{s}$,
$A_{s}$, $\xi_{1}$, $\xi_{2}$ and $\epsilon$. We altered the CosmoMC
to include the new parameters $\xi_{1}$, $\xi_{2}$ and $\epsilon$.
To find the best-fit values for the parameters, we use the recent
five-year WMAP (WMAP5) dataset \cite{WMAPdata}. We find that our
model gives a better fit with $\Delta\chi^{2}\sim 8.5$ as compared
to the standard inflationary model without trans-Planckian
corrections. This improvement in the value of WMAP $\chi^{2}$ has
been obtained because of the difference in the $\xi_{1}$ and
$\xi_{2}$ coefficients. We also checked that there is no
improvement in $\chi^{2}$ in the Danielsson model with equal
coefficients $\xi_{1}=\xi_{2}$. In Fig. (\ref{fig1}) we have
plotted the 1D marginalized posterior probability and the
normalized mean likelihood for all of the primordial parameters of
our model. In Fig. (\ref{fig2}) the contours of the 2D
marginalized posterior probability and the normalized mean
likelihoods are displayed for each parameters pairs. The shading
shows the mean likelihood of the samples and helps to demonstrate
when the marginalized probability is enhanced by a longer
parameter space rather than by a better fit to the data. Using
CAMB \cite{lewis2} we have plotted the TT and TE angular power
spectrum in Fig. (\ref{fig3}) and the matter power spectrum in
Fig. (\ref{fig5}), corresponding to our trans-Planckian model. The
presence of small oscillations in the trans-Planckian angular TT
power spectrum for $l<200$, is the reason that we obtain a smaller
value for the likelihood $\chi^{2}$.
\begin{table*}
\begin{tabular}{|r|c|c|c|c|c|c|c|c|c|c|c|}
\hline $ \xi_{1}$ & $\xi_{2}\times 10^{4}$ & $h$ & $\Omega_{b}
h^2$ &$\Omega_{c}h^2 $ & $\Omega_{\Lambda}$ & $\tau$& $A_{s}\times 10^{10}$ & $\epsilon$ & $n_{s}$ & $\chi^{2}$/d.o.f. \\
\hline $0$ & $-$ & $0.71$ & $0.023$ & $0.11$ & $0.73$ & $0.083$ &
$21.32 $ & $-$ & $0.96$ & $2658.48/1453$\\
$0.309$ & $8.14 $ & $0.72$ & $0.022$ & $0.11$ & $0.75$ & $0.089$ &
$21.03$ & $0.0798$ & $0.95$ & $2649.96/1456$\\
\hline
\end{tabular}
\caption{Best fit parameters from the WMAP data for the standard
model of inflation compared to the best fit parameters obtained
for our trans-Planckian model. }\label{bestfit}
\end{table*}


Table \ref{bestfit} shows the best fit parameters found by CosmoMC
for the case $\xi_{1}=0$ (standard inflationary model) and
$\xi_{1}\neq 0$ (trans-Planckian model). For the case of
$\xi_{1}=0$ the standard values for the cosmological
parameters \cite{standard} are recovered.
We can estimate the scale
of noncommutativity, $\frac{1}{\sqrt{\theta}}=\Lambda_{NC}$, from
the trans-Planckian parameter $\xi_{2}=H/\Lambda_{NC}\approx
10^{-4}$. If the scale of $H$ during inflation is approximated to
be of the order of $H\geq 10^{12}$ GeV \cite{nong}, then the
noncommutative scale is estimated as $\Lambda_{NC}\geq 10^{16} $
GeV. On the other hand as Refs.
\cite{allah1,allah2,allah3,allah4} have demonstrated, the Minimal
Supersymmetric Standard Model (MSSM) has all the ingredients to
give rise to successful inflation. The MSSM inflation occurs at
low scales i.e. $H\geq 1$ GeV. Then this put a bound on the scale of
noncommutativity as $\Lambda_{NC}\geq 10^{4}$ GeV. Similar bounds on noncommutative scale,
has reported in \cite{hyd2} using the investigation of Lorentz symmetry violation due to
noncommutativity during experiments and also in \cite{hyd1} by studying of
hydrogen atom spectrum and the Lamb-shift effect on a noncommutative space which is $\Lambda_{NC}\geq 10$ TeV.

%%%%%%%%%%%%%%%%%%%%%%%%%%%%%%%%%%%%%%%%%%%%%%%%%%%%%%%%%%%%%%%%%%%%%%%%%%%%%%%%%%%%%%%%%%%%%%%%%%%%%%%%%%%%
\section{\large Conclusion}

We showed that a modification in the action of one inflaton
due to noncommutativity, leads to a nonzero Bogoliubov
coefficient $\beta_{k}$ which affects the power spectrum to have
oscillatory corrections.  Using CosmoMC code we showed that these
corrections cause a drop in the WMAP $\chi^{2}$ of about
$\Delta\chi^{2}\sim 8.5$. We also derived the noncommutative scale
using parameters estimated by CosmoMC. Tacking the scale of
inflation to be $H\geq 10^{12}$ GeV, we can estimate the scale of
noncommutativity which yields $\Lambda_{NC}\geq 10^{16}$ GeV and
with the scale of inflation $H\geq 1$ GeV, we get $\Lambda_{NC}\geq
10^{4}$ GeV.

\section*{\small Acknowledgment}

We would like to thank F.~Loran for reading the manuscript and for
very useful discussion and suggestions.

%%%%%%%%%%%%%%%%%%%%%%%%%%%%%%%%%%%%%%%%%%%%%%%%%%%%%%%%%%%%%%%%%%%%%%%%%%%%%%%%%%


\begin{thebibliography}{99}


\bibitem{brand1}
J.~Martin and R.~H.~Brandenberger, {\it The Trans-Planckian problem of inflationary cosmology},
 Phys. Rev. D{\bf 63} (2001) 123501, [hep-th/0005209].

\bibitem{brand2}
J.~Martin and R.~H.~Brandenberger, {\it A cosmological window on trans-Planckian physics
}, [astro-ph/0012031].

\bibitem{brand3}
S.~Tsujikawa, R.~Maartens and R.~Brandenberger, {\it Non-commutative inflation and the CMB },
Phys. Lett. B{\bf 574} (2003) 141, [astro-ph/0308169].

\bibitem{greene1}
R.~Easther, B.~R.~Greene, W.~H.~Kinney and G.~Shiu, {\it Imprints of short distance
physics on inflationary cosmology }, Phys. Rev. D{\bf 67} (2003) 063508, [hep-th/0110226];   S. Shankaranarayanan, {\it Is there an imprint of Planck scale physics on inflationary cosmology?}, Class. Quant. Grav. {\bf 20} (2003) 75,
[gr-qc/0203060]; A.~Ashoorioon, J.~L.~Hovdebo and R.~B.~Mann,
   {\it Running of the spectral index and violation of the consistency relation
   between tensor and scalar spectra from trans-Planckian physics},
   Nucl.\ Phys.\  B {\bf 727}, 63 (2005),
   [arXiv:gr-qc/0504135].

\bibitem{greene2}
R.~Easther, B.~R.~Greene, W.~H.~Kinney and G.~Shiu, {\it  A generic estimate of trans-Planckian
modifications to the primordial power spectrum in inflation },  Phys. Rev. D{\bf 66} (2002) 023518, [hep-th/0204129].

\bibitem{greene3}
B.~Greene, M.~Parikh and J.~P.~van der Schaar, {\it Universal correction to the inflationary vacuum},
JHEP {\bf 0604} (2006) 057, [hep-th/0512243].

\bibitem{tanaka}
T.~Tanaka, {\it  A comment on trans-Planckian physics in inflationary universe }, [astro-ph/0012431].

\bibitem{niemeyer1}
 J.~C.~Niemeyer and R.~Parentani, {\it Trans-Planckian dispersion and scale-invariance of inflationary perturbations
}, Phys. Rev. D{\bf 64} (2001) 101301, [astro-ph/0101451].

\bibitem{niemeyer2}
 A.~Kempf and J.~C.~Niemeyer, {\it  Perturbation spectrum in inflation with cutoff }, Phys. Rev. D{\bf 64} (2001) 103501,
 [astro-ph/0103225].

\bibitem{niemeyer3}
D.~Campo, J.~Niemeyer and R.~Parentani, {\it Damped corrections to inflationary spectra from a fluctuating cutoff },
arXiv:0705.0747v1 [hep-th].


\bibitem{starob}
A.~A.~Starobinsky, {\it Robustness of the inflationary perturbation spectrum to trans-Planckian physics },
Pisma Zh. Eksp. Teor. Fiz. {\bf 73} (2001) 415; JETP Lett. {\bf 73} (2001) 371, [astro-ph/0104043].

\bibitem{daniel1}
 U.~H.~Danielsson, {\it A note on inflation and transplanckian physics }, Phys. Rev. D{\bf 66} (2002) 023511,
 [hep-th/0203198].

\bibitem{low}
K.~Goldstein and D.~A.~Lowe, {\it Initial state effects on the cosmic microwave background and trans-planckian physics },
Phys. Rev. D{\bf 67} (2003) 063502, [hep-th/0208167].

\bibitem{burgess}
C.~P.~Burgess, J.~M.~Cline, F.~Lemieux and R.~Holman, {\it  Are inflationary predictions
sensitive to very high energy physics?}, JHEP {\bf 0302} (2003) 048, [hep-th/0210233].


\bibitem{paren}
J.~Macher and R.~Parentani, {\it Signatures of trans-Planckian dispersion in inflationary spectra},
arXiv:0804.1920v3 [hep-th].


\bibitem{anderson}
P.~R.~Anderson, C.~Molina-Paris and E.~Mottola, {\it  Short distance and initial state effects
in inflation: stress tensor and decoherence },  Phys. Rev. D{\bf 72} (2005) 043515, [hep-th/0504134].


\bibitem{lim}
C.~Armendariz-Picon and E.~A.~Lim, {\it Vacuum choices and the predictions of inflation },
JCAP {\bf 0312} (2003) 006, [hep-th/0303103].

\bibitem{holman1}
H.~Collins and R.~Holman, {\it  Trans-planckian signals from the breaking of local Lorentz invariance },
arXiv: 0705.4666v1 [hep-ph].


\bibitem{book1}
A.~R.~Liddle and D.~H.~Lyth, {\it Cosmological inflation and large-scale structure}, Cambridge
University Press (2000).

\bibitem{book2}
V.~Mukhanov, {\it Physical foundations of cosmology}, Cambridge University Press (2005).

\bibitem{burgess2}
C.~P.~Burgess, {\it  Lectures on cosmic inflation and its potential stringy realizations
}, arXiv: 0708.2865v1 [hep-th].

\bibitem{linde}
A.~Linde, {\it Inflationary cosmology}, arXiv: 0705.0164v2 [hep-th].

\bibitem{garcia}
J.~Garcia-Bellido, {\it Cosmology and astrophysics }, [astro-ph/0502139].


\bibitem{niemeyer4}
J.~Niemeyer, R.~Parentani and D.~Campo, {\it Minimal modifications of the primordial
power spectrum from an adiabatic short distance cutoff  },
arXiv: 0705.0747v1 [hep-th].

\bibitem{brand4}
J.~Martin and R.~H.~Brandenberger, {\it  On the dependence of
the spectra of fluctuations in inflationary cosmology on trans-Planckian physics },
Phys. Rev. D{\bf 68} (2003) 063513, [hep-th/0305161].

\bibitem{kaloper}
N.~Kaloper, M.~Kleban, A.~E.~Lawrence and S.~ Shenker, {\it Signatures of short distance
physics in the Cosmic Microwave Background }
Phys. Rev. D {\bf 66}, 123510 (2002) [hep-th/0201158].

\bibitem{energy1}
R.~H.~Brandenberger and J.~Martin, {\it Back-Reaction and the
Trans-Planckian problem of inflation revisited }, Phys. Rev. D{\bf
71} (2005) 023504, [hep-th/0410223]; M.~Lemoine, M.~Lubo, J.~Martin and J.~P.~Uzan,
{\it The stress-energy tensor for trans-Planckian cosmology}, Phys.\ Rev.\  D {\bf 65} (2002) 023510,
[arXiv:hep-th/0109128].

\bibitem{brand5}
R.~H.~Brandenberger and J.~Martin, {\it On Signatures of short distance physics in the Cosmic Microwave Background },
Int. J. Mod. Phys. A{\bf17} (2002) 3663, [hep-th/0202142].

\bibitem{weinberg}
S.~Weinberg, {\it Effective field theory for inflation}, arXiv:0804.4291v2 [hep-th].

\bibitem{WMAP1}
J.~Martin and C.~Ringeval, {\it superimposed oscillations in the
WMAP data?}, Phys. Rev. D{\bf 69} (2004) 083515,
[astro-ph/0310382].

\bibitem{WMAP2}
J.~Martin and C.~Ringeval, {\it Addendum to "Superimposed
oscillations in the WMAP data?" }, Phys. Rev. D{\bf 69} (2004)
127303, [astro-ph/0402609].

\bibitem{WMAP3}
J.~Martin and C.~Ringeval, {\it Exploring the Superimposed
Oscillations Parameter Space}, JCAP {\bf 0501} (2005) 007,
[hep-ph/0405249].

\bibitem{WMAP11}
 N.~E.~Groeneboom and O.~Elgaroy, {\it Detection of transplanckian effects in the cosmic microwave
 background}, Phys. Rev. D {\bf 77},
043522 (2008), arXiv: 0711.1793v4 [astro-ph]

\bibitem{book3}
N.~D.~Birrell and P.~C.~W.~Davies, {\it Quantum fields in curved space}, Cambridge University
Press, 1982.

\bibitem{BD}
T.~S.~Bunch and P.~C.~Davies, Proc. Roy. Soc. Lond. A{\bf 360}
(1978) 117.

\bibitem{shalm}
K.~Schalm, G.~Shiu, J.~P.~ van der Schaar, {\it The cosmological vacuum ambiguity, effective
actions, and transplanckian effects in inflation}, [hep-th/0412288].

\bibitem{book4}
I.~S.~Gradshteyn and I.~M.~Ryzhik, {\it Table of Integrals, Series, and Products
}, Fifth edition, Academic Press Inc., London, (1994).

\bibitem{pad}
L. Sriramkumar and  T. Padmanabhan, {\it Initial state of matter fields and
trans-Planckian physics: Can CMB observations disentangle the two?} Phys.
Rev. D{\bf 71} (2005) 103512,  [gr-qc/0408034].

\bibitem{cai1}
Yi-Fu Cai, Tao-tao Qiu, Jun-Qing Xia, Xinmin Zhang, {\it
A Model Of Inflationary Cosmology Without Singularity},
arXiv:0808.0819 [astro-ph]. 

\bibitem{cai2}
Yi-fu Cai and Yun-Song Piao, {\it
Probing noncommutativity with inflationary gravitational waves},
Phys. Lett. B{\bf 657} (2007) 1, [gr-qc/0701114].

\bibitem{2}
M.~R.~douglas and N.~A.~Nekrasov, {\it Noncommutative field
theory}, Rev.\ Mod.\ Phys.\ {\bf 73} (2001) 977, [hep-th/0106048].

\bibitem{3}
R.~J.~Szabo, {\it Quantum field theory on noncommutative space},
Phys.\ Rept.\ {\bf 378} (2003) 207, [hep-th/0109162].

\bibitem{douglas}
M.~R.~douglas and N.~A.~Nekrasov, {\it Noncommutative field theory}, Rev. Mod. Phys.
{\bf 73} (2001) 977, [hep-th/0106048].

\bibitem{szabo}
R.~J.~Szabo, {\it Quantum field theory on noncommutative space}, Phys. Rept. {\bf 378}
(2003) 207, [hep-th/0109162].

\bibitem{Minwalla}
S.~Minwalla, M.~V.~Raamsdonk and N.~Seiberg, {\it Noncommutative perturbative dynamics},
JHEP {\bf 02} (2000) 020, [hep-th/9912072].

\bibitem{grosse}
H.~Grosse and R.~Wulkenhaar, {\it Renormalisation of $\phi^4$-theory on noncommutative $R^4$ in the matrix base
}, Commun. Math. Phys. {\bf 256} (2005) 305, [hep-th/0401128].

\bibitem{gurau}
R.~Gurau, J.~Magnen, V.~Rivasseau and A.~Tanasa, {\it A translation-invariant renormalizable non-commutative scalar model
}, arXiv: 0802.0791v1 [math-ph].

\bibitem{khodam}
B.~Mirza, M.~Zarei, {\it Effective field theory of a locally noncommutative space-time and extra dimensions
}, arXiv: 0803.0232v1 [hep-th].

\bibitem{szabo2}
E.~Langmann, R.~J.~Szabo, {\it Duality in Scalar Field Theory on Noncommutative
Phase Spaces}, Phys. Lett. B{\bf 533} (2001) 168, [hep-th/0202039].

\bibitem{lewis1}
A.~Lewis and S.~Bridle, Phys. Rev. D{\bf 66}, 103511 (2002),
[astro-ph/0205436], http://cosmologist.info/cosmomc.

\bibitem{daniel2}
L.~Bergstrom and U~.H.~Danielsson, {\it Can MAP and Planck map
Planck physics?},  JHEP {\bf 0212} (2002) 038, [hep-th/0211006].

\bibitem{hu}
W.~Hu and S.~Dodelson, {\it Cosmic Microwave Background
Anisotropies}, Ann. Rev. Astron. Astrophys. {\bf 40} (2002) 171,
[astro-ph/0110414].

\bibitem{WMAP4}
J.~M.~Cline, P.~Crotty and J.~Lesgourgues, {\it Does the small CMB
quadrupole moment suggest new physics? }, JCAP {\bf 0309} (2003)
010, [astro-ph/0304558].

\bibitem{WMAP5}
S.~Hannestad and L.~Mersini-Houghton, {\it A first glimpse of
string theory in the sky? }, Phys. Rev. D{\bf 71} (2005) 123504,
[hep-ph/0405218].

\bibitem{WMAP6}
C.~R.~Contaldi, M.~Peloso, L.~Kofman and A.~Linde, {\it
Suppressing the lower Multipoles in the CMB Anisotropies }, JCAP
{\bf 0307} (2003) 002, [astro-ph/0303636].

\bibitem{WMAP7}
R.~Easther, W.~H.~Kinney and H.~Peiris, {\it Boundary Effective
Field Theory and Trans-Planckian Perturbations: Astrophysical
Implications}, JCAP {\bf 0508} (2005) 001, [astro-ph/0505426].

\bibitem{WMAP8}
N.~E.~Groeneboom and O.~Elgaroy, {\it  Detection of transplanckian
effects in the cosmic microwave background}, Phys. Rev. D{\bf 77}
043522 (2008), arXiv:0711.1793v4 [astro-ph].

\bibitem{WMAP9}
O.~Elgaroy and S.~Hannestad, {\it Can Planck-scale physics be seen
in the cosmic microwave background ?}, Phys.Rev. D{\bf 68} (2003)
123513, [astro-ph/0307011].

\bibitem{WMAP10}
J.~Hamann, S.~Hannestad, M.~S.~Sloth and Y.~Y.~Y.~Wong, {\it
Observing trans-Planckian ripples in the primordial power spectrum
with future large scale structure probes}, arXiv: 0807.4528v1
[astro-ph].

\bibitem{lewis2}
A. Lewis, A. Challinor and A. Lasenby, Astrophys. J. {\bf 538},
473 (2000), [astro-ph/9911177], http://camb.info.

\bibitem{WMAPdata}
http://lambda.gsfc.nasa.gov

\bibitem{standard}
E. Komatsu et al., {\it Five-Year Wilkinson Microwave Anisotropy
Probe (WMAP) Observations: Cosmological Interpretation},
 arXiv:0803.0547 [astro-ph].

\bibitem{nong}
N. Bartolo, E. Komatsu, S. Matarrese and A. Riotto, {\it
Non-Gaussianity from Inflation: Theory and Observations}, Phys.
Rept. {\bf 402} (2004) 103, [astro-ph/0406398].

\bibitem{allah1}
R. Allahverdi, J. Garcia-Bellido, K. Enqvist and A. Mazumdar, {\it
Gauge invariant MSSM inflaton}, Phys. Rev. Lett. {\bf 97} (2006)
191304, [hep-ph/0605035].

\bibitem{allah2}
R. Allahverdi, A. Kusenko and A. Mazumdar, {\it A-term inflation
and the smallness of the neutrino masses}, [hep-ph/0608138].


\bibitem{allah3}
R. Allahverdi, K. Enqvist, J. Garcia-Bellido, A. Jokinen and A.
Mazumdar, {\it MSSM flat direction inflation: slow roll,
stability, fine tunning and reheating}, [hep-ph/0610134].

\bibitem{allah4}
R. Allahverdi, B. Dutta and A. Mazumdar, {\it Unifying inflation
and dark matter with neutrino masses }, arXiv: 0708.3983 [hep-ph].

\bibitem{hyd2}
S.~M.~Carroll, J.~A.~Harvey, V.~A.~Kostelecky, C.~D.~Lane,
T.~Okamoto, {\it Noncommutative Field Theory and Lorentz
Violation}, Phys. Rev. Lett. {\bf 87} (2001) 141601,
[hep-th/0105082].

\bibitem{hyd1}
M. Chaichian, M. M. Sheikh-Jabbari, A. Tureanu, {\it Hydrogen atom
spectrum and the Lamb shift in noncommutative QED}, Phys. Rev.
Lett. {\bf 86} (2001) 2716, [hep-th/0010175].

\bibitem{2dF}
 S.~Cole and et al., {\it The 2dF Galaxy Redshift Survey: Power-spectrum analysis of
the final dataset and cosmological implications} Mon. Not. Roy.
Astron. Soc. {\bf 362} (2005) 505, [astro-ph/0501174].


%\bibitem{martin3}
%M.~Lemoine, J.~Martin and J.P.~Uzan, {\it Trans-Planckian dark
%energy?},  Phys. Rev. D{\bf 67} (2003) 103520, [hep-th/0212027].




\end{thebibliography}
\end{document}